\documentclass[aps,pre,showpacs]{revtex4}
\usepackage{amssymb,amsmath}
\usepackage[dvips]{graphicx}

\newcommand{\bd}[1]{\mbox{\boldmath$#1$}}

\newcommand{\Pd}[1]{\ensuremath{\Bigl[\!\!\Bigl[#1\Bigr]\!\!\Bigr]}}

\begin{document}

\title{Collective phase description of globally coupled excitable elements}

\author{Yoji Kawamura}
\email{ykawamura@jamstec.go.jp}
\affiliation{Institute for Research on Earth Evolution,
Japan Agency for Marine-Earth Science and Technology, Yokohama 236-0001, Japan}

\author{Hiroya Nakao}
%\email{nakao@mei.titech.ac.jp}
%\email{nakao@ton.scphys.kyoto-u.ac.jp}
%\affiliation{Department of Physics, Graduate School of Science, Kyoto University, Kyoto 606-8502, Japan}
%\affiliation{Department of Physics, Kyoto University, Kyoto 606-8502, Japan}
%\affiliation{Department of Mechanical and Environmental Informatics, Tokyo Institute of Technology,
%2-12-1 Ookayama, Meguro, Tokyo 152-8552, Japan}
\affiliation{Department of Mechanical and Environmental Informatics, Tokyo Institute of Technology, Tokyo 152-8552, Japan}
%\affiliation{CREST, Japan Science and Technology Agency, Kawaguchi 332-0012, Japan}
\affiliation{JST, CREST, Kyoto 606-8502, Japan}

%\author{Kensuke Arai}
%\email{karai@brain.riken.jp}
%\affiliation{Brain Science Institute, RIKEN, Wako 351-0198, Japan}

%\author{Hiroshi Kori}
%\email{kori.hiroshi@ocha.ac.jp}
%\affiliation{Division of Advanced Sciences, Ochadai Academic Production, Ochanomizu University, Tokyo 112-8610, Japan}
%\affiliation{PRESTO, Japan Science and Technology Agency, Kawaguchi 332-0012, Japan}

\author{Yoshiki Kuramoto}
%\email{kuramoto@kurims.kyoto-u.ac.jp}
\affiliation{Research Institute for Mathematical Sciences, Kyoto University, Kyoto 606-8502, Japan}
\affiliation{Institute for Integrated Cell-Material Sciences, Kyoto University, Kyoto 606-8501, Japan}

%\date{\today}
%\date{January 1, 2011}
%\date{June 18, 2011}  % submitted to Phys. Rev. E
\date{October 5, 2011} % submitted to arXiv

%% 05.45.-a Nonlinear dynamics and chaos
%% 05.45.Xt Synchronization; coupled oscillators
%% 82.40.-g Chemical kinetics and reactions: special regimes and techniques
%% 82.40.Bj Oscillations, chaos, and bifurcations
%% 82.40.Ck Pattern formation in reactions with diffusion, flow and heat transfer

%%%%% abstract
\begin{abstract}
  We develop a theory of collective phase description
  for globally coupled noisy excitable elements exhibiting macroscopic oscillations.
  Collective phase equations describing macroscopic rhythms of the system
  are derived from Langevin-type equations of globally coupled active rotators
  via a nonlinear Fokker-Planck equation.
  The theory is an extension of the conventional phase reduction method for ordinary limit cycles
  to limit-cycle solutions in infinite-dimensional dynamical systems,
  such as the time-periodic solutions to nonlinear Fokker-Planck equations
  representing macroscopic rhythms.
  We demonstrate that
  the type of the collective phase sensitivity function near the onset of collective oscillations
  crucially depends on the type of the bifurcation,
  namely, it is type-I for the saddle-node bifurcation and type-II for the Hopf bifurcation.
\end{abstract}

\maketitle

%%%%% section 1
\section{Introduction} \label{sec:introduction}

%synchronization~\cite{ref:winfree80,ref:kuramoto84,ref:pikovsky01,ref:strogatz03,ref:manrubia04}.
%review of active rotators~\cite{ref:lindner04,ref:acebron05,ref:sagues07}.
%active rotators~\cite{ref:shinomoto86,ref:sakaguchi88,ref:kurrer95,ref:tanabe99,ref:kanamaru03,ref:zaks03,ref:tessone07}.
%interacting groups~\cite{ref:okuda91,ref:montbrio04}.
%Ott-Antonsen ansatz~\cite{ref:ott08,ref:ott09,ref:ott11}.
%collective phase (i)~\cite{ref:kawamura07,ref:kawamura08,ref:kawamura10a}.
%collective phase (iii)~\cite{ref:kawamura10b}.
%collective phase (ii)~\cite{ref:kori09}.
%type-I and type-II~\cite{ref:hansel95,ref:ermentrout96}.
%Schwabedal-Pikovsky~\cite{ref:schwabedal10a,ref:schwabedal10b}.
%Langevin-type equation~\cite{ref:risken89,ref:gardiner97}.
%Malkin theorem~\cite{ref:hoppensteadt97,ref:izhikevich07}.
%Nakao paper~\cite{ref:nakao11}. \\

The collective rhythms emerging from coupled dynamical elements
are the most remarkable synchronization phenomena in nature
~\cite{ref:winfree80,ref:kuramoto84,ref:pikovsky01,ref:strogatz03,ref:manrubia04}.
These collective rhythms, or collective oscillations, are exhibited
not only by populations of self-oscillatory elements,
but also by populations of excitable elements,
in which the elemental dynamics are not self-oscillatory
but their coupling gives rise to nontrivial collective dynamics
~\cite{ref:lindner04,ref:acebron05,ref:sagues07}.
The active rotator is a well-known, convenient phenomenological model
that can describe both excitable and oscillatory dynamics.
Systems of interacting active rotators have been extensively studied
and their collective dynamics have been analyzed
~\cite{ref:shinomoto86,ref:sakaguchi88,ref:kurrer95,ref:tanabe99,ref:kanamaru03,ref:zaks03,ref:tessone07}.

Macroscopic synchronization between interacting groups
of globally coupled phase oscillators exhibiting collective rhythms has attracted our attention,
because many rhythms in the real world are actually collective rhythms
generated by coupled networks of microscopic elements~\cite{ref:okuda91,ref:montbrio04}.
The recent introduction of the Ott-Antonsen ansatz~\cite{ref:ott08,ref:ott09,ref:ott11}
has facilitated theoretical investigations in this direction,
although applicable only to a restricted class of coupled phase oscillators.
As an alternative general approach
to study the phase synchronization between macroscopic rhythms,
we have formulated a theory of collective phase description for coupled phase oscillators,
which gives a reduced collective phase equation for the macroscopic rhythms
~\cite{ref:kawamura07,ref:kawamura08,ref:kawamura10a,ref:kawamura10b,ref:kori09}.
With this method,
we can analyze the synchronization properties between collective rhythms
in a closed way at the macroscopic level.
Each group of the oscillators can be treated as a single macroscopic oscillator,
which facilitates theoretical investigations of interacting collective rhythms.
In particular, we have derived a formula
that relates the collective phase response of the macroscopic rhythm
to the individual phase responses of the constituent oscillators
~\cite{ref:kawamura07,ref:kawamura08,ref:kawamura10a,ref:kawamura10b,ref:kori09}.

In this study,
we develop a more generalized theory of collective phase description
that is applicable to macroscopic rhythms
generated by excitable elements as well as by oscillatory elements.
Specifically, we consider interacting groups of globally coupled noisy active rotators
and derive a coupled collective phase equation describing them.
Our present theory can be applied to
a broader class of interacting groups of globally coupled dynamical elements,
and includes our previous theory for globally coupled noisy phase oscillators
~\cite{ref:kawamura07,ref:kawamura08,ref:kawamura10a} as a special case.
As an example,
we analyze the type of the collective phase sensitivity function near the onset of collective oscillations,
and reveal that it crucially depends on the type of bifurcation,
as in the phase sensitivity functions of ordinary dynamical systems~\cite{ref:hansel95,ref:ermentrout96}.

Recently, Schwabedal and Pikovsky~\cite{ref:schwabedal10a,ref:schwabedal10b}
studied somewhat similar issues on effective phase description
of noise-induced oscillations in excitable systems.
While their study treated a one-body problem
for an ensemble of statistically independent excitable elements at the microscopic level,
the present study focuses on globally coupled noisy excitable elements exhibiting collective oscillations,
which can be considered a one-body problem at the macroscopic level.
Specifically, the focus of this study is on the development of a phase reduction method
for genuinely collective oscillations arising from mutual interactions in coupled dynamical systems.

The organization of the present paper is as follows.
In Sec.~\ref{sec:formulation},
we formulate a theory of collective phase description
for globally coupled noisy excitable elements.
In Sec.~\ref{sec:simulation}, we illustrate our theory using numerical simulations
and reveal that the collective phase sensitivity function behaves very differently,
depending on the bifurcation type of the collective oscillations.
Concluding remarks are given in Sec.~\ref{sec:conclusion}.
The attached appendices clarify the relation of our present theory
to the conventional phase reduction theory of ordinary limit-cycle oscillators (App.~\ref{sec:A}),
present the results for a system of globally coupled phase oscillators with translational symmetry (App.~\ref{sec:B}),
and outline further extensions to general systems of globally coupled dynamical elements (App.~\ref{sec:C}).

%%%%% section 2
\section{Formulation of collective phase description} \label{sec:formulation}

In this section, we formulate a theory of collective phase description
for globally coupled noisy active rotators exhibiting macroscopic oscillations.
The theory can be considered a phase reduction method
for limit-cycle solutions in infinite-dimensional dynamical systems.
The relation of the present theory to the conventional phase reduction
of ordinary limit-cycle solutions is given in App.~\ref{sec:A}.
The present theory is also an extension (which we explain in App.~\ref{sec:B})
of our previous theory for phase oscillators~\cite{ref:kawamura07,ref:kawamura08,ref:kawamura10a}.

%%% subsection
\subsection{Langevin-type equations and the nonlinear Fokker-Planck equation}

We consider interacting groups of globally coupled active rotators described by the following equation:
%%% eq
\begin{equation}
  \dot{\phi}_j^{(\sigma)}\left( t \right)
  = v\left( \phi_j^{(\sigma)} \right)
  + \frac{1}{N} \sum_{k=1}^N \Gamma\left( \phi_j^{(\sigma)}, \phi_k^{(\sigma)} \right)
  + \sqrt{D}\, \xi_j^{(\sigma)}\left( t \right)
  + \epsilon_p Z\left( \phi_j^{(\sigma)} \right) p_\sigma\left( t \right)
  + \frac{\epsilon_g}{N} \sum_{k=1}^N \Gamma_{\sigma\tau}\left( \phi_j^{(\sigma)}, \phi_k^{(\tau)} \right),
  \label{eq:model}
\end{equation}
for $j = 1, \cdots, N$ and $(\sigma, \tau) = (1, 2)$ or $(2, 1)$,
where $\phi_j^{(\sigma)}(t)$ is the state of the $j$-th active rotator at time $t$
in the $\sigma$-th group consisting of $N$ elements.
The first term on the right-hand side represents the intrinsic dynamics of the active rotator;
the second term, the internal coupling between the elements in the same group;
the third term, the noise independently applied to each rotator;
the fourth term, the external forcing common to all rotators in the $\sigma$-th group;
and the last term, the external coupling between the elements belonging to different groups.
Each isolated active rotator obeys $\dot\phi = v(\phi)$,
which describes either excitable or oscillatory dynamics~\cite{ref:shinomoto86}.
The internal and external coupling functions are given by
$\Gamma(\phi_j^{(\sigma)}, \phi_k^{(\sigma)})$ and
$\Gamma_{\sigma\tau}(\phi_j^{(\sigma)}, \phi_k^{(\tau)})$, respectively.
The characteristic intensity of the internal coupling within a group is scaled to unity,
whereas that of the external coupling between the groups is given by $\epsilon_g \geq 0$.
The state-dependent sensitivity function of the active rotator is $Z(\phi_j^{(\sigma)})$,
which corresponds to the phase sensitivity function in the case of phase oscillators.
We assume that the rotator variable $\phi$ takes values in $[0, 2\pi]$
and that the above functions, i.e.,
$v(\phi)$, $Z(\phi)$, $\Gamma(\phi, \phi')$, and $\Gamma_{\sigma\tau}(\phi, \phi')$,
are $2\pi$-periodic in the respective state variables.
The external forcing is denoted by $p_\sigma(t)$,
whose characteristic intensity is given by $\epsilon_p \geq 0$.
The noise $\xi_j^{(\sigma)}(t)$
is assumed to be independent white Gaussian noise~\cite{ref:risken89,ref:gardiner97},
whose statistics are given by
%%% eq
\begin{equation}
  \left\langle \xi_j^{(\sigma)}\left( t \right) \right\rangle = 0, \qquad
  \left\langle \xi_j^{(\sigma)}\left( t \right) \xi_k^{(\tau)}\left( s \right) \right\rangle
  = 2 \delta_{jk} \delta_{\sigma\tau} \delta\left( t - s \right).
\end{equation}
The noise intensity is characterized by $D \geq 0$.

In the continuum limit, i.e., $N \to \infty$,
the Langevin-type equation~(\ref{eq:model}) describing coupled groups of active rotators
can be transformed into the coupled nonlinear Fokker-Planck equation as follows
(see also Refs.~\cite{ref:kuramoto84,ref:acebron05,ref:shinomoto86,ref:okuda91,ref:kawamura07,ref:kawamura10a}).
Let $f^{(\sigma)}(\phi, t)$ represent the one-body probability density function
of rotator state $\phi$ in the $\sigma$-th group, which is normalized as
%%% eq
\begin{equation}
  \int_0^{2\pi} d\phi\, f^{(\sigma)}\left( \phi, t \right) = 1.
\end{equation}
Because the mean-field theory exactly holds
for the case of global coupling in the continuum limit~\cite{ref:kuramoto84},
we can average the internal and external coupling terms in Eq.~(\ref{eq:model})
by the one-body probability density functions,
$f^{(\sigma)}(\phi, t)$ and $f^{(\tau)}(\phi, t)$, respectively.
Then the Langevin-type equation~(\ref{eq:model})
takes the form of a single-rotator equation as follows:
%%% eq
\begin{equation}
  \dot{\phi}_j^{(\sigma)}\left( t \right)
  = V^{(\sigma)}\left( \phi_j^{(\sigma)}, t \right)
  + \sqrt{D}\, \xi_j^{(\sigma)}\left( t \right),
  \label{eq:single-rotator}
\end{equation}
where
%%% eq
\begin{align}
  V^{(\sigma)}\left( \phi_j^{(\sigma)}, t \right)
  =&\,\, v\left( \phi_j^{(\sigma)} \right)
  + \int_0^{2\pi} d\phi' \, \Gamma\left( \phi_j^{(\sigma)}, \phi' \right)
  f^{(\sigma)}\left( \phi', t \right) \nonumber \\
  &+ \epsilon_p Z\left( \phi_j^{(\sigma)} \right) p_\sigma\left( t \right)
  + \epsilon_g \int_0^{2\pi} d\phi' \, \Gamma_{\sigma\tau}\left( \phi_j^{(\sigma)}, \phi' \right)
  f^{(\tau)}\left( \phi', t \right).
\end{align}
The single-rotator Langevin-type equation~(\ref{eq:single-rotator})
can be transformed into the following coupled nonlinear Fokker-Planck equation:
%%% eq
\begin{equation}
  \frac{\partial}{\partial t} f^{(\sigma)}\left( \phi, t \right) =
  -\frac{\partial}{\partial \phi} \left[ V^{(\sigma)}\left( \phi, t \right) f^{(\sigma)}\left( \phi, t \right) \right]
  + D \frac{\partial^2}{\partial \phi^2} f^{(\sigma)}\left( \phi, t \right),
\end{equation}
or, explicitly,
%%% eq
\begin{align}
  \frac{\partial}{\partial t} f^{(\sigma)}\left( \phi, t \right) =
  &-\frac{\partial}{\partial \phi} \left[ \left\{
    v\left( \phi \right) + \int_0^{2\pi} d\phi'\, \Gamma\left( \phi, \phi' \right) f^{(\sigma)}\left( \phi', t \right)
    \right\} f^{(\sigma)}\left( \phi, t \right) \right]
  + D \frac{\partial^2}{\partial \phi^2} f^{(\sigma)}\left( \phi, t \right)
  \nonumber \\
  &- \epsilon_p \frac{\partial}{\partial \phi} \left[
    Z\left( \phi \right) f^{(\sigma)}\left( \phi, t \right) \right] p_\sigma\left( t \right)
  - \epsilon_g \frac{\partial}{\partial \phi} \left[
    \int_0^{2\pi} d\phi'\, \Gamma_{\sigma\tau}\left( \phi, \phi' \right) f^{(\tau)}\left( \phi', t \right)
    f^{(\sigma)}\left( \phi, t \right) \right].
  \label{eq:nfp}
\end{align}
The first two terms on the right-hand side of the nonlinear Fokker-Planck equation~(\ref{eq:nfp})
represent the internal dynamics of the $\sigma$-th group,
the third term represents the external forcing applied to the $\sigma$-th group,
and the last term represents the external coupling between the $\sigma$-th group and the $\tau$-th group.
When the external forcing and the external coupling are absent,
i.e., when $\epsilon_p = \epsilon_g = 0$,
each group of active rotators obeying Eq.~(\ref{eq:nfp})
is assumed to exhibit stable collective oscillation, i.e., a stable time-periodic solution.
We further assume that this situation persists
even if $\epsilon_p$ and/or $\epsilon_g$ become slightly positive.

%%% subsection
\subsection{Collectively oscillating solution and its Floquet eigenfunctions}

Let us assume $\epsilon_p = \epsilon_g = 0$ and focus on a single group.
The group index $\sigma$ is dropped in this subsection.
Then the nonlinear Fokker-Planck equation~(\ref{eq:nfp}) with $\epsilon_p = \epsilon_g = 0$
can be written in the following form:
%%% eq
\begin{equation}
  \frac{\partial}{\partial t} f\left( \phi, t \right) =
  -\frac{\partial}{\partial \phi} \left[ \left\{
    v\left( \phi \right) + \int_0^{2\pi} d\phi'\, \Gamma\left( \phi, \phi' \right) f\left( \phi', t \right)
    \right\} f\left( \phi, t \right) \right]
  + D \frac{\partial^2}{\partial \phi^2} f\left( \phi, t \right).
  \label{eq:single-nfp}
\end{equation}
We now assume the existence of a stable time-periodic solution
to the nonlinear Fokker-Planck equation~(\ref{eq:single-nfp}).
It is known that such a solution exists
in a certain model of globally coupled active rotators~\cite{ref:shinomoto86},
which we analyze in Sec.~\ref{sec:simulation}.
The stable time-periodic solution
to the nonlinear Fokker-Planck equation~(\ref{eq:single-nfp}) can be described by
%%% eq
\begin{equation}
  f\left( \phi, t \right) = f_0\left( \phi, \Theta(t) \right), \qquad
  \dot{\Theta}(t) = \Omega,
  \label{eq:solution}
\end{equation}
where $\Theta$ and $\Omega$ are the collective phase and collective frequency, respectively.
Inserting Eq.~(\ref{eq:solution}) into the nonlinear Fokker-Planck equation~(\ref{eq:single-nfp}),
we find that $f_0(\phi, \Theta)$ satisfies the following equation:
%%% eq
\begin{equation}
  \Omega \frac{\partial}{\partial \Theta} f_0\left( \phi, \Theta \right) =
  -\frac{\partial}{\partial \phi} \left[ \left\{
    v\left( \phi \right) + \int_0^{2\pi} d\phi'\, \Gamma\left( \phi, \phi' \right) f_0\left( \phi', \Theta \right)
    \right\} f_0\left( \phi, \Theta \right) \right]
  + D\frac{\partial^2}{\partial \phi^2} f_0\left( \phi, \Theta \right).
  \label{eq:f0}
\end{equation}
Let $u(\phi, \Theta, t)$ represent a small disturbance to the unperturbed collectively oscillating solution,
and we consider a slightly perturbed solution
%%% eq
\begin{equation}
  f\left( \phi, t \right) = f_0\left( \phi, \Theta(t) \right) + u\left( \phi, \Theta(t), t \right).
\end{equation}
Equation~(\ref{eq:single-nfp}) is then linearized in $u(\phi, \Theta, t)$, i.e.,
%%% eq
\begin{equation}
  \frac{\partial}{\partial t} u\left( \phi, \Theta, t \right)
  = \hat{L}\left( \phi, \Theta \right) u\left( \phi, \Theta, t \right).
  \label{eq:linear}
\end{equation}
Here, the linear operator $\hat{L}(\phi, \Theta)$ is given by
%%% eq
\begin{equation}
  \hat{L}\left( \phi, \Theta \right) u\left( \phi, \Theta \right)
  = \left[ \hat{J}\left( \phi, \Theta \right) - \Omega \frac{\partial}{\partial \Theta} \right]
  u\left( \phi, \Theta \right),
\end{equation}
where
%%% eq
\begin{align}
  \hat{J}\left( \phi, \Theta \right) u\left( \phi, \Theta \right) =
  &-\frac{\partial}{\partial \phi} \Bigl[ v\left( \phi \right) u\left( \phi, \Theta \right) \Bigr]
  -\frac{\partial}{\partial \phi} \left[ u\left( \phi, \Theta \right)
    \int_0^{2\pi} d\phi'\, \Gamma\left( \phi, \phi' \right) f_0\left( \phi', \Theta \right) \right] \nonumber \\
  &-\frac{\partial}{\partial \phi} \left[ f_0\left( \phi, \Theta \right)
    \int_0^{2\pi} d\phi'\, \Gamma\left( \phi, \phi' \right) u\left( \phi', \Theta \right) \right]
  + D\frac{\partial^2}{\partial \phi^2} u\left( \phi, \Theta \right).
\end{align}
Note that $\hat{L}(\phi, \Theta)$ is time-periodic through $\Theta$,
and therefore, Eq.~(\ref{eq:linear}) is a Floquet-type system with a periodic linear operator.
Defining the inner product of $2\pi$-periodic functions as
%%% eq
\begin{equation}
  \Pd{ u^\ast\left( \phi, \Theta \right),\, u\left( \phi, \Theta \right) }
  = \frac{1}{2\pi} \int_0^{2\pi} d\Theta \int_0^{2\pi} d\phi\,
  u^\ast\left( \phi, \Theta \right) u\left( \phi, \Theta \right),
  \label{eq:inner}
\end{equation}
we introduce an adjoint operator of $\hat{L}(\phi, \Theta)$ by
%%% eq
\begin{equation}
  \Pd{ u^\ast\left( \phi, \Theta \right),\,
    \hat{L}\left( \phi, \Theta \right) u\left( \phi, \Theta \right) }
  = \Pd{ \hat{L}^\ast\left( \phi, \Theta \right) u^\ast\left( \phi, \Theta \right),\,
    u\left( \phi, \Theta \right) }.
\end{equation}
By partial integration, the adjoint operator $\hat{L}^\ast(\phi, \Theta)$ is explicitly given by
%%% eq
\begin{equation}
  \hat{L}^\ast\left( \phi, \Theta \right) u^\ast\left( \phi, \Theta \right)
  = \left[ \hat{J}^\ast\left( \phi, \Theta \right) + \Omega \frac{\partial}{\partial \Theta} \right]
  u^\ast\left( \phi, \Theta \right),
\end{equation}
where
%%% eq
\begin{align}
  \hat{J}^{\ast}\left( \phi, \Theta \right) u^\ast\left( \phi, \Theta \right)
  =&\,\, v\left( \phi \right) \frac{\partial}{\partial \phi} u^\ast\left( \phi, \Theta \right)
  + \left[ \int_0^{2\pi} d\phi'\, \Gamma\left( \phi, \phi' \right) f_0\left( \phi', \Theta \right) \right]
  \frac{\partial}{\partial \phi} u^\ast\left( \phi, \Theta \right) \nonumber \\
  &+ \int_0^{2\pi} d\phi'\, \Gamma\left( \phi', \phi \right) f_0\left( \phi', \Theta \right)
  \frac{\partial}{\partial \phi'} u^\ast\left( \phi', \Theta \right)
  + D\frac{\partial^2}{\partial \phi^2} u^\ast\left( \phi, \Theta \right).
\end{align}
In the calculation below,
we use the Floquet eigenfunctions of the periodic linear system Eq.~(\ref{eq:linear}) and its adjoint system
associated with the zero eigenvalue, i.e.,
%%% eq
\begin{align}
  \hat{L}\left( \phi, \Theta \right) u_0\left( \phi, \Theta \right)
  = \left[ \hat{J}\left( \phi, \Theta \right) - \Omega \frac{\partial}{\partial \Theta} \right]
  u_0\left( \phi, \Theta \right)
  &= 0, \\
  \hat{L}^\ast\left( \phi, \Theta \right) u_0^\ast\left( \phi, \Theta \right)
  = \left[ \hat{J}^\ast\left( \phi, \Theta \right) + \Omega \frac{\partial}{\partial \Theta} \right]
  u_0^\ast\left( \phi, \Theta \right)
  &= 0. \label{eq:leftzero}
\end{align}
Note that the right zero eigenfunction $u_0(\phi, \Theta)$ can be chosen as
%%% eq
\begin{equation}
  u_0\left( \phi, \Theta \right) = \frac{\partial}{\partial \Theta} f_0\left( \phi, \Theta \right),
  \label{eq:u0}
\end{equation}
as is confirmed by differentiating Eq.~(\ref{eq:f0}) with respect to $\Theta$.
Using the inner product~(\ref{eq:inner}) with the right zero eigenfunction~(\ref{eq:u0}),
the left zero eigenfunction $u_0^\ast(\phi, \Theta)$ is normalized as
%%% eq
\begin{equation}
  \Pd{ u_0^\ast\left( \phi, \Theta \right),\, u_0\left( \phi, \Theta \right) }
  = \frac{1}{2\pi} \int_0^{2\pi} d\Theta \int_0^{2\pi} d\phi\,
  u_0^\ast\left( \phi, \Theta \right) u_0\left( \phi, \Theta \right)
  = 1.
\end{equation}
Here, we should note that the following equation holds
(see also Eq.~(\ref{eq:secA-dtheta}) and Ref.~\cite{ref:hoppensteadt97})
%%% eq
\begin{align}
  \frac{\partial}{\partial\Theta}
  \left[ \int_0^{2\pi} d\phi\, u_0^\ast\left( \phi, \Theta \right) u_0\left( \phi, \Theta \right) \right]
  &= \int_0^{2\pi} d\phi\, \left[ u_0^\ast\left( \phi, \Theta \right)
  \frac{\partial}{\partial\Theta} u_0\left( \phi, \Theta \right)
  + u_0\left( \phi, \Theta \right)
  \frac{\partial}{\partial\Theta} u_0^\ast\left( \phi, \Theta \right) \right]
  \nonumber \\
  &= \frac{1}{\Omega} \int_0^{2\pi} d\phi\, \left[ u_0^\ast\left( \phi, \Theta \right)
  \hat{J}\left( \phi, \Theta \right) u_0\left( \phi, \Theta \right)
  - u_0\left( \phi, \Theta \right)
  \hat{J}^\ast\left( \phi, \Theta \right) u_0^\ast\left( \phi, \Theta \right) \right]
  \nonumber \\
  &= 0.
\end{align}
Therefore, the following normalization condition is satisfied independently for each $\Theta$ as
%%% eq
\begin{equation}
  \int_0^{2\pi} d\phi\, u_0^\ast\left( \phi, \Theta \right) u_0\left( \phi, \Theta \right) = 1.
\end{equation}
In the following subsections,
using the collectively oscillating solution and its Floquet zero eigenfunctions,
we formulate a theory of collective phase description for globally coupled active rotators.

%%% subsection
\subsection{Collective phase reduction}

Let us introduce the effects of weak external forcing and external coupling,
i.e., we assume $0 < \epsilon_p \ll 1$ and $0 < \epsilon_g \ll 1$.
We treat the third term and the last term in Eq.~(\ref{eq:nfp}) as perturbations.
Using the idea of phase reduction~\cite{ref:kuramoto84},
we can derive a coupled collective phase equation
from the nonlinear Fokker-Planck equation~(\ref{eq:nfp}).
Namely, we project the dynamics of the nonlinear Fokker-Planck equation~(\ref{eq:nfp})
onto the unperturbed collectively oscillating solution as
%%% eq
\begin{align}
  \dot \Theta^{(\sigma)}(t)
  =& \int_0^{2\pi} d\phi\, u_0^\ast\left( \phi, \Theta^{(\sigma)} \right)
  \frac{\partial}{\partial t} f^{(\sigma)}\left( \phi, t \right)
  \nonumber \\
  \simeq& \, \Omega
  - \epsilon_p \int_0^{2\pi} d\phi\, u_0^\ast\left( \phi, \Theta^{(\sigma)} \right)
  \frac{\partial}{\partial \phi} \left[
    Z\left( \phi \right) f_0\left( \phi, \Theta^{(\sigma)} \right) \right] p_\sigma\left( t \right)
  \nonumber \\
  &- \epsilon_g \int_0^{2\pi} d\phi\, u_0^\ast\left( \phi, \Theta^{(\sigma)} \right)
  \frac{\partial}{\partial \phi} \left[
    \int_0^{2\pi} d\phi'\, \Gamma_{\sigma\tau}\left( \phi, \phi' \right)
    f_0\left( \phi', \Theta^{(\tau)} \right)
    f_0\left( \phi, \Theta^{(\sigma)} \right) \right],
\end{align}
where we approximate $f^{(\sigma)}(\phi,t)$
by the unperturbed solution $f_0(\phi,\Theta^{(\sigma)})$
and use the fact that
%%% eq
\begin{equation}
  \int_0^{2\pi} d\phi\, u_0^\ast\left( \phi, \Theta \right)
  \frac{\partial}{\partial t} f_0\left( \phi, \Theta \right)
  = \Omega \int_0^{2\pi} d\phi\, u_0^\ast\left( \phi, \Theta \right)
  u_0\left( \phi, \Theta \right)
  = \Omega.
\end{equation}
Thus, the collective phase equation is obtained in the following form:
%%% eq
\begin{equation}
  \dot{\Theta}^{(\sigma)}\left( t \right) = \Omega
  + \epsilon_p \zeta\left( \Theta^{(\sigma)} \right) p_\sigma\left( t \right)
  + \epsilon_g \digamma_{\sigma\tau}\left( \Theta^{(\sigma)}, \Theta^{(\tau)} \right).
  \label{eq:cpr}
\end{equation}
Here, the {\it collective phase sensitivity function} is given by
%%% eq
\begin{equation}
  \zeta\left( \Theta^{(\sigma)} \right)
  = \int_0^{2\pi} d\phi\, Z\left( \phi \right) k_0\left( \phi, \Theta^{(\sigma)} \right),
  \label{eq:zeta}
\end{equation}
and the {\it effective phase coupling function} is given by
%%% eq
\begin{equation}
  \digamma_{\sigma\tau}\left( \Theta^{(\sigma)}, \Theta^{(\tau)} \right)
  = \int_0^{2\pi} d\phi \int_0^{2\pi} d\phi'\, \Gamma_{\sigma\tau}\left( \phi, \phi' \right)
  k_0\left( \phi, \Theta^{(\sigma)} \right) f_0\left( \phi', \Theta^{(\tau)} \right),
  \label{eq:digamma}
\end{equation}
where the kernel function is defined as
%%% eq
\begin{equation}
  k_0\left( \phi, \Theta \right)
  = f_0\left( \phi, \Theta \right)
  \frac{\partial}{\partial \phi} u_0^\ast\left( \phi, \Theta \right).
\end{equation}
In the next subsection,
we simplify the collective phase equation
by applying the near-identity transformation.

%%% subsection
\subsection{Normal form of the collective phase equation}

Applying the near-identity transformation~\cite{ref:kuramoto84,ref:hoppensteadt97,ref:kori09} to Eq.~(\ref{eq:cpr}),
namely, by slightly (i.e., on the order of $O(\epsilon_g)$) changing the definition of the collective phase $\Theta$,
we can obtain the following collective phase equation
%%% eq
\begin{equation}
  \dot{\Theta}^{(\sigma)}(t) = \Omega
  + \epsilon_p \zeta\left( \Theta^{(\sigma)} \right) p_\sigma\left( t \right)
  + \epsilon_g \gamma_{\sigma\tau}\left( \Theta^{(\sigma)} - \Theta^{(\tau)} \right),
  \label{eq:cpe}
\end{equation}
where the {\it collective phase coupling function} is given by
%%% eq
\begin{equation}
  \gamma_{\sigma\tau}\left( \Theta^{(\sigma)} - \Theta^{(\tau)} \right)
  = \frac{1}{2\pi} \int_0^{2\pi} d\lambda\,
  \digamma_{\sigma\tau}\left( \lambda + \Theta^{(\sigma)}, \lambda + \Theta^{(\tau)} \right),
  \label{eq:average}
\end{equation}
or, more explicitly, by
%%% eq
\begin{equation}
  \gamma_{\sigma\tau}\left( \Theta \right)
  = \frac{1}{2\pi} \int_0^{2\pi} d\lambda \int_0^{2\pi} d\phi \int_0^{2\pi} d\phi'\,
  \Gamma_{\sigma\tau}\left( \phi, \phi' \right)
  k_0\left( \phi, \lambda + \Theta \right) f_0\left( \phi', \lambda \right).
  \label{eq:gamma}
\end{equation}
Note that the derivation of Eq.~(\ref{eq:cpe}) from Eq.~(\ref{eq:cpr})
follows procedures developed in the appendix of Ref.~\cite{ref:kori09}
where the details of the near-identity transformation are given.
In addition, when the external forcing is absent, i.e., when $\epsilon_p = 0$,
Eq.~(\ref{eq:average}) can be understood as the result of the averaging method~\cite{ref:kuramoto84}.
In summary, the collective phase equation~(\ref{eq:cpe})
with the collective phase sensitivity and the collective phase coupling functions,
i.e., Eq.~(\ref{eq:zeta}) and Eq.~(\ref{eq:gamma}),
are derived from the Langevin-type equation~(\ref{eq:model})
via the nonlinear Fokker-Planck equation~(\ref{eq:nfp}).

%%% subsection
\subsection{Adjoint method for the left zero eigenfunction}

Finally, we explain a numerical method for obtaining the left zero eigenfunction.
From Eq.~(\ref{eq:leftzero}),
the left zero eigenfunction $u_0^\ast(\phi,\Theta)$ satisfies
%%% eq
\begin{equation}
  \Omega \frac{\partial}{\partial\Theta} u_0^\ast\left( \phi, \Theta \right)
  = - \hat{J}^\ast\left( \phi, \Theta \right) u_0^\ast\left( \phi, \Theta \right),
\end{equation}
which can be transformed into
%%% eq
\begin{equation}
  \frac{\partial}{\partial s} u_0^\ast\left( \phi, -\Omega s \right)
  = \hat{J}^\ast\left( \phi, -\Omega s \right) u_0^\ast\left( \phi, -\Omega s \right)
  \label{eq:adjoint}
\end{equation}
with $\Theta = -\Omega s$.
A relaxation method using Eq.~(\ref{eq:adjoint}),
which numerically calculates the eigenfunction associated with the zero eigenvalue
by evolving Eq.~(\ref{eq:adjoint})
until all eigenfunctions with negative eigenvalues decay,
is a convenient method for obtaining the left zero eigenfunction
(see also Eq.~(\ref{eq:secA-adjoint}) and
Refs.~\cite{ref:ermentrout96,ref:hoppensteadt97,ref:izhikevich07}). \\

As mentioned at the beginning of this section,
in App.~\ref{sec:A}, we clarify the relation between our present formulation
and the conventional phase reduction of ordinary limit-cycle solutions.
In particular, it should be noted that
the variable $\phi$ of the nonlinear Fokker-Planck equation
plays the role of the vector component index in App.~\ref{sec:A}.
In App.~\ref{sec:B}, it is shown that the present formulation
includes our previous theory for phase oscillators~\cite{ref:kawamura07,ref:kawamura08,ref:kawamura10a}
as a special case with continuous translational symmetry in $\phi$.

%%%%% section 3
\section{Numerical analysis of globally coupled active rotators} \label{sec:simulation}

In this section, we illustrate our theory (developed in the preceding section)
using numerical simulations.
First, we introduce a complex order parameter and its related quantities,
which we will use to characterize the collective oscillations.
Thereafter, we present the results of numerical simulations.
In particular, we analyze the type of the collective phase sensitivity function
near the onset of collective oscillations.

%%% subsection
\subsection{Order parameters}

To quantify the collective oscillations,
we introduce a complex order parameter~\cite{ref:kuramoto84}
%%% eq
\begin{equation}
  R(t) e^{i\Phi(t)}
  = \frac{1}{N} \sum_{k=1}^N e^{i\phi_k(t)}
  = \int_0^{2\pi} d\phi\, e^{i\phi} f\left(\phi, t \right),
  \label{eq:A}
\end{equation}
which is time-dependent.
We also use the following real constant order parameter~\cite{ref:shinomoto86}
%%% eq
\begin{equation}
  S = \left\langle \Bigl| R e^{i\Phi} - \left\langle R e^{i\Phi} \right\rangle_t \Bigr| \right\rangle_t,
  \label{eq:S}
\end{equation}
where $\langle \cdot \rangle_t$ is a long-time average.
The order parameter $S$ is non-zero only when collective oscillations exist.
In addition, using the real and imaginary parts of the complex order parameter,
we introduce the following pair of real order parameters:
%%% eq
\begin{equation}
  X = R \cos \Phi, \qquad
  Y = R \sin \Phi.
  \label{eq:XY}
\end{equation}
In visualizing the limit-cycle orbit of the nonlinear Fokker-Planck equation
with the infinite-dimensional state space,
we project the collectively oscillating solution onto the $X$-$Y$ plane as
%%% eq
\begin{equation}
  X_0(\Theta) + i Y_0(\Theta)
  = R_0(\Theta) e^{i\Phi_0(\Theta)}
  = \int_0^{2\pi} d\phi\, e^{i\phi} f_0\left(\phi, \Theta \right).
\end{equation}
%%

%%% subsection
\subsection{Collectively oscillating solution and other associated functions}

In numerical simulations,
we use the following model of globally coupled active rotators~\cite{ref:shinomoto86}:
%%% eq
\begin{equation}
  v(\phi) = \omega - r \sin(\phi), \qquad
  \omega, r \geq 0,
\end{equation}
and
%%% eq
\begin{equation}
  \Gamma\left( \phi, \phi' \right)
  = \Gamma_{\sigma\tau}\left( \phi, \phi' \right)
  = -\sin\left( \phi - \phi' \right).
\end{equation}
For simplicity,
the state-dependent sensitivity function of the active rotator is assumed to be constant, i.e.,
%%% eq
\begin{equation}
  Z(\phi) = 1.
\end{equation}
Here, we summarize several properties of the above model~\cite{ref:shinomoto86}:
(i) The active rotators are excitable when $\omega < r$,
whereas they are oscillatory when $\omega > r$.
(ii) In the noiseless case, i.e., $D = 0$,
the model exhibits collective oscillations under the condition $\omega > r$,
which emerges via a saddle-node bifurcation.
(iii) In the phase-oscillator limit, i.e., $r = 0$,
the model exhibits collective oscillations under the condition $D < D_{\rm c} = 1/2$,
which occurs via a supercritical Hopf bifurcation.

In our numerical simulations,
the parameters of the active rotators are fixed at $\omega = 0.50$ and $r = 0.52$,
values that satisfy the excitable condition $\omega < r$.
The noise intensity is mostly chosen to be $D = 0.22$,
for which collective oscillations exist
with the collective frequency $\Omega \simeq 0.1955$.
In numerical simulations of the nonlinear Fokker-Planck equation~(\ref{eq:nfp})
and the adjoint equation~(\ref{eq:adjoint}),
we used the pseudospectral method with $2^{10}$ modes.

\begin{figure*}
  \begin{center}
    \includegraphics[width=0.9\hsize,clip]{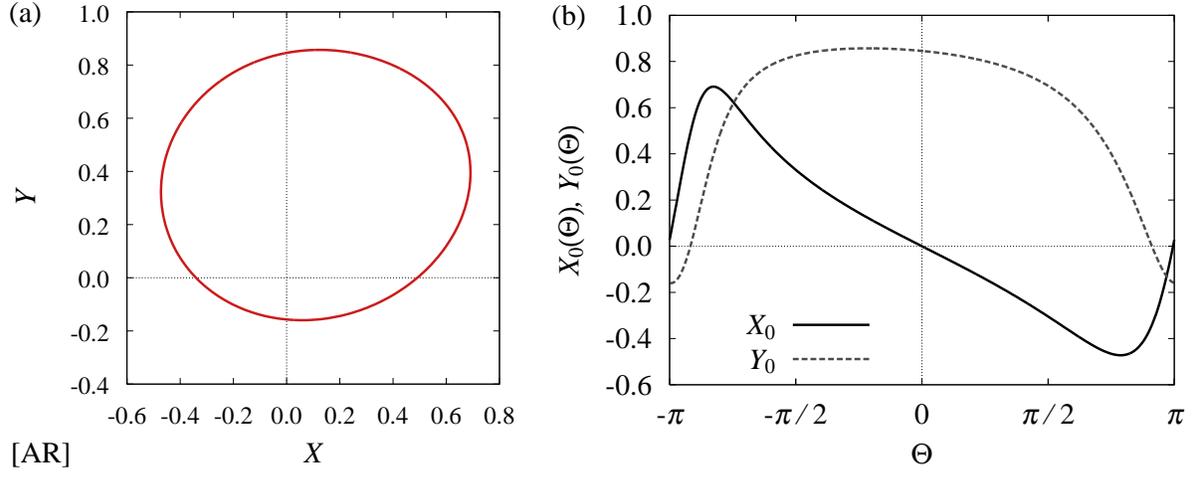}
    \caption{(Color online)
      Globally coupled active rotators (AR) in the excitable regime exhibiting macroscopic rhythms.
      (a) Macroscopic limit-cycle orbit projected on the $X$-$Y$ plane.
      (b) Wave forms of $X_0(\Theta)$ and $Y_0(\Theta)$.
      Parameters are $\omega = 0.50$, $r = 0.52$, and $D = 0.22$,
      which give $\Omega \simeq 0.1955$.
    }
    \label{fig:1}
  \end{center}
  %%%%% parameter
  %% 
  %% v(\phi) = \omega - r \sin(\phi),
  %% \Gamma(\phi, \phi') = -\sin(\phi - \phi').
  %% 
  %% \omega = 0.50, r = 0.52, D = 0.22.
  %% 
  %% stationary state: phi_0 = asin(\omega/r) = 1.2925495.
  %% [ reference value1:               2\pi/5 = 1.25663706. ]
  %% [ reference value2:                \pi/2 = 1.57079633. ]
  %% 
  %% collective frequency: \Omega = 0.1955.
  %% 
  %%%%%
\end{figure*}

\begin{figure*}
  \begin{center}
    \includegraphics[width=0.9\hsize,clip]{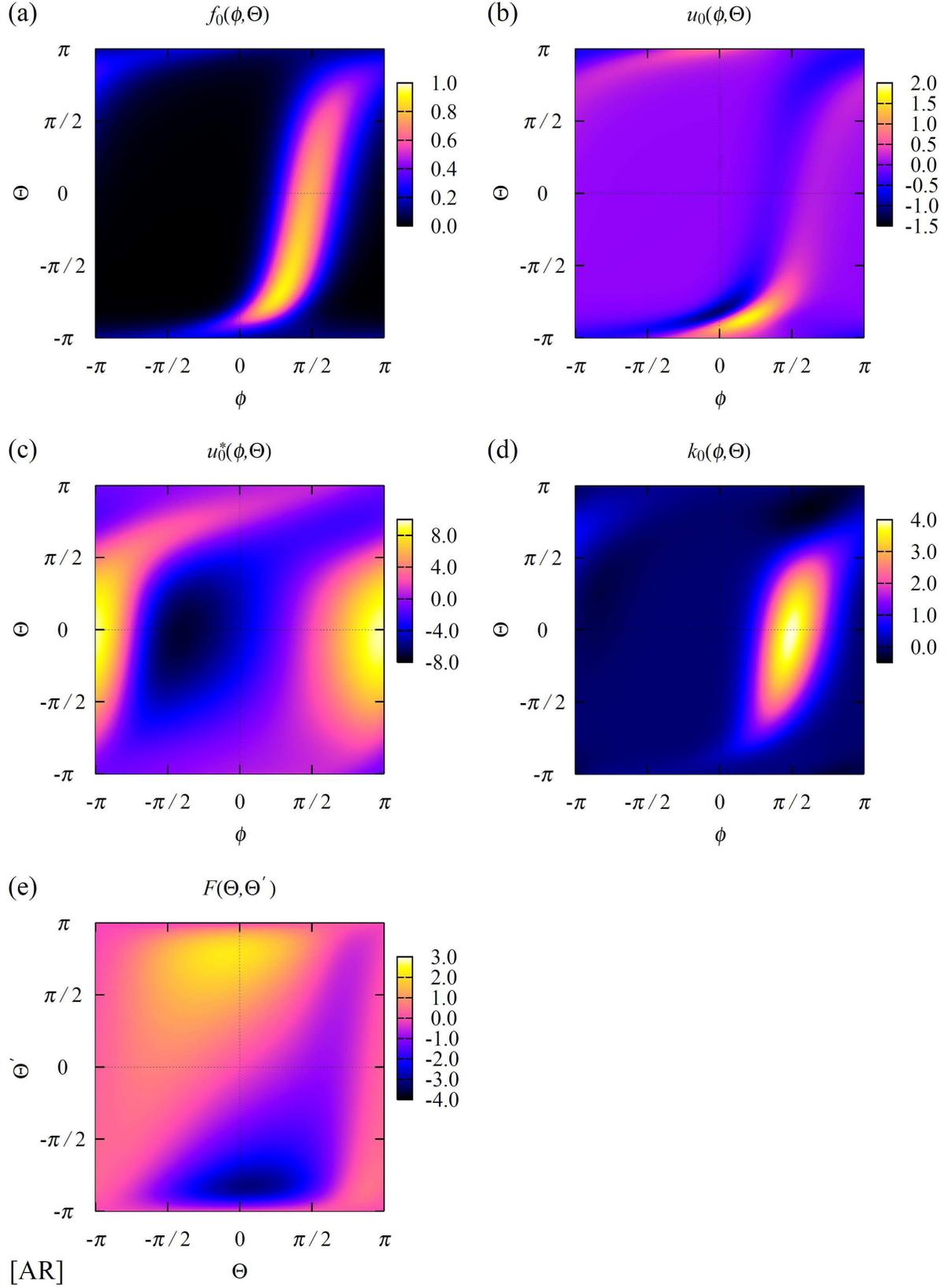}
    \caption{(Color online)
      Globally coupled active rotators (AR).
      (a) Collectively oscillating solution $f_0(\phi, \Theta)$.
      (b) Right zero eigenfunction $u_0(\phi, \Theta)$.
      (c) Left zero eigenfunction $u_0^\ast(\phi, \Theta)$.
      (d) Kernel function $k_0(\phi, \Theta)$.
      (e) Effective phase coupling function $\digamma(\Theta, \Theta')$.
      Parameters are $\omega = 0.50$, $r = 0.52$, and $D = 0.22$.
    }
    \label{fig:2}
  \end{center}
  %%%%% parameter
  %% 
  %% mode number of \phi   is 1024.
  %% mode number of \Theta is 1024.
  %% 
  %% other parameters: see Fig.1.
  %% 
  %%%%%
\end{figure*}

Figure~\ref{fig:1} shows the limit-cycle orbit projected on the $X$-$Y$ plane,
which was obtained from the nonlinear Fokker-Planck equation~(\ref{eq:nfp})
with $\epsilon_p = \epsilon_g = 0$, i.e., Eq.~(\ref{eq:single-nfp}).
The collectively oscillating solution $f_0(\phi,\Theta)$ and other associated functions,
i.e., $u_0(\phi,\Theta)$, $u_0^\ast(\phi,\Theta)$, $k_0(\phi,\Theta)$, and $\digamma(\Theta,\Theta')$,
are displayed in Fig.~\ref{fig:2}.

\begin{figure*}
  \begin{center}
    \includegraphics[width=0.9\hsize,clip]{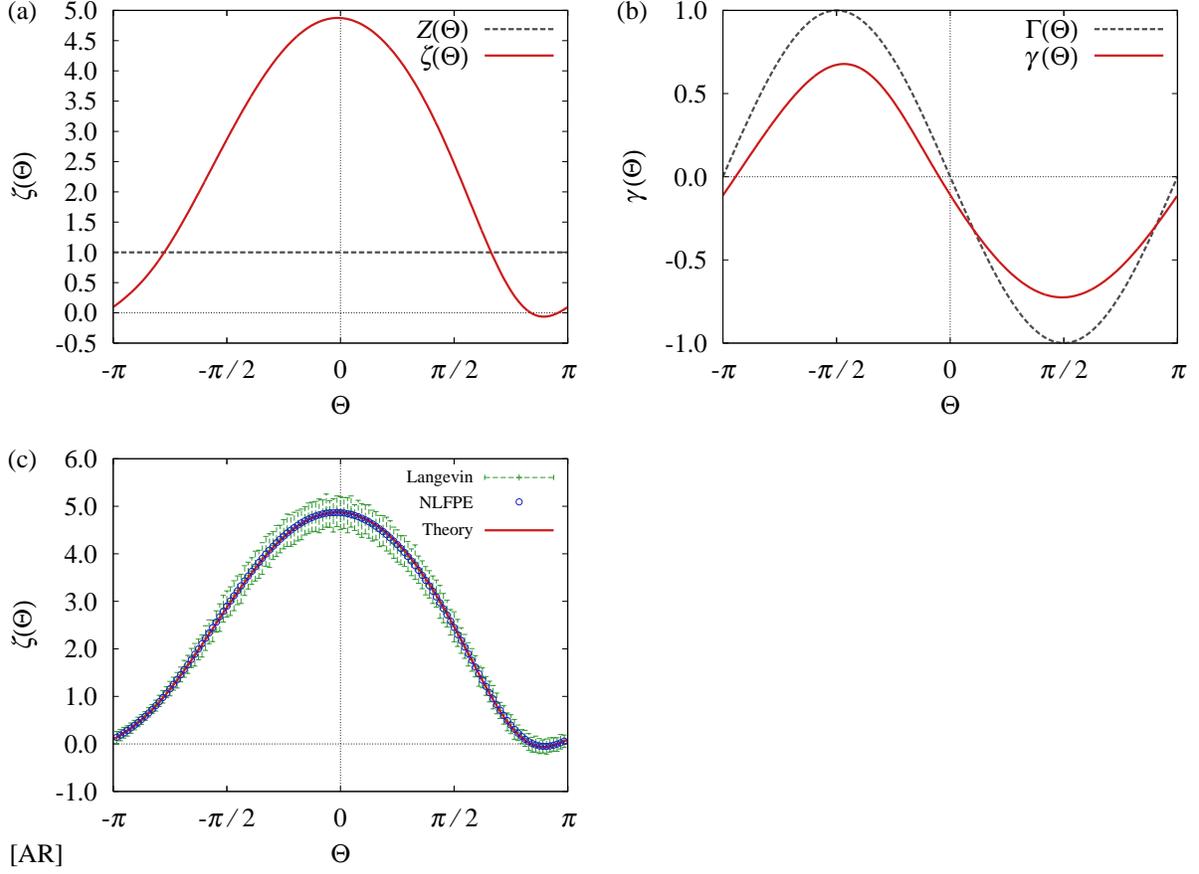}
    \caption{(Color online)
      Globally coupled active rotators (AR).
      (a) Collective phase sensitivity function $\zeta(\Theta)$.
      (b) Collective phase coupling function $\gamma(\Theta)$.
      (c) Comparison of collective phase sensitivity functions $\zeta(\Theta)$
      between theoretical curve (Theory) and direct numerical simulations with impulse intensity $\epsilon_p = 0.01$,
      i.e., Langevin-type equation (Langevin) and nonlinear Fokker-Planck equation (NLFPE).
      Parameters are $\omega = 0.50$, $r = 0.52$, and $D = 0.22$.
      In Langevin-type simulations, the number of active rotators in the system is $N = 10^5$
      and the results are averaged over $100$ ensembles,
      whose means and standard deviations are shown.
    }
    \label{fig:3}
  \end{center}
  %%%%% parameter (a)(b)
  %% 
  %% microscopic state-dependent sensitivity function: Z(\phi) = 1.
  %% microscopic phase coupling function: \Gamma(\phi, \phi') = -sin(\phi - \phi').
  %% 
  %% other parameters: see Figs.1 and 2.
  %% 
  %%%%% parameter (c)
  %% 
  %% \epsilon_p = 0.01.
  %% 
  %% [NLFPE]
  %% mode number of \phi is 1024.
  %% 
  %% [Langevin]
  %% number of rotators is 10^5.
  %% ensemble number is 100.
  %% 
  %% other parameters: see Fig.1.
  %% 
  %%%%%
\end{figure*}

The collective phase sensitivity function $\zeta(\Theta)$
and the collective phase coupling function $\gamma_{\sigma\tau}(\Theta)$
are shown in Fig.~\ref{fig:3}(a) and Fig.~\ref{fig:3}(b), respectively.
The collective phase sensitivity function $\zeta(\Theta)$
is quite different from the individual state-dependent sensitivity function $Z(\phi) = 1$.
Despite the fact that $Z(\phi)$ is constant,
the collective phase sensitivity function $\zeta(\Theta)$ depends on $\Theta$,
in sharp contrast to the case of globally coupled phase oscillators
where $\zeta(\Theta)$ never depends on $\Theta$ if $Z(\phi)$ is constant~\cite{ref:kawamura08}
(see also App.~\ref{sec:B}).
Consequently, although the external coupling function has only a first harmonic,
i.e., $\Gamma_{\sigma\tau}(\phi,\phi') = -\sin(\phi-\phi')$,
the collective phase coupling function $\gamma_{\sigma\tau}(\Theta)$ includes higher harmonics (and also a uniform component);
that is also in contrast with the case of globally coupled phase oscillators
in which $\gamma_{\sigma\tau}(\Theta)$ is sinusoidal whenever $\Gamma_{\sigma\tau}(\phi)$ is sinusoidal~\cite{ref:kawamura10a}
(see also App.~\ref{sec:B}).

The collective phase sensitivity function $\zeta(\Theta)$,
measured by direct numerical simulations of the system
by applying sufficiently weak stimulating impulses with $\epsilon_p = 0.01$, % and $\epsilon_g = 0$,
is compared to the theoretical curve in Fig.~\ref{fig:3}(c).
The results of the nonlinear Fokker-Planck equation
corresponding to the continuum limit $N \to \infty$ agree perfectly with our theory;
the results of the Langevin-type simulation
with $N = 10^5$ active rotators averaged over $100$ ensembles
agree with our theory within the fluctuations due to finite size effects.
Therefore, the formulation developed in Sec.~\ref{sec:formulation}
has been validated by the above-described numerical simulations.

%%% subsection
\subsection{Types of collective phase sensitivity function}

Now, we investigate the dependence of the collective phase sensitivity function $\zeta(\Theta)$
on the noise intensity $D$.
In particular, we analyze the type of the collective phase sensitivity function $\zeta(\Theta)$
near the onset of collective oscillations.

\begin{figure*}
  \begin{center}
    \includegraphics[width=0.9\hsize,clip]{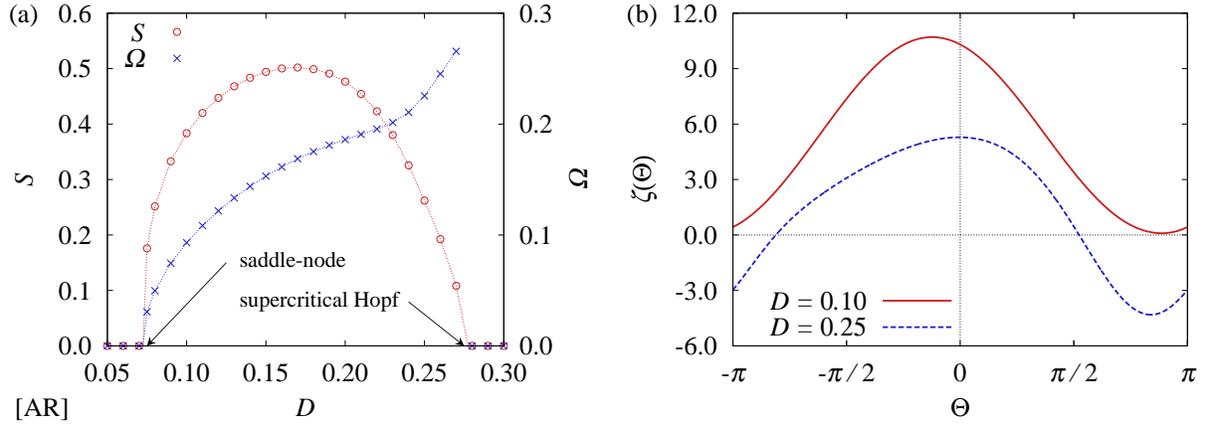}
    \caption{(Color online)
      Globally coupled active rotators (AR).
      (a) Dependence of order parameter $S$ and collective frequency $\Omega$
      on noise intensity $D$.
      (b) Collective phase sensitivity function $\zeta(\Theta)$.
      Parameters are $\omega = 0.50$ and $r = 0.52$.
      The collective phase sensitivity function near the saddle-node bifurcation is type-I,
      whereas that near the Hopf bifurcation is type-II.
    }
    \label{fig:4}
  \end{center}
  %%%%% parameter
  %% 
  %% v(\phi) = \omega - r \sin(\phi),
  %% \Gamma(\phi, \phi') = -\sin(\phi - \phi').
  %% \omega = 0.50, r = 0.52.
  %% 
  %%%%%
\end{figure*}

Figure~\ref{fig:4}(a) shows
the dependence of both the order parameter $S$ and the collective frequency $\Omega$
on the noise intensity $D$ (see also Ref.~\cite{ref:shinomoto86}).
As the the noise intensity is increased,
collective oscillations arise via a saddle-node bifurcation
and disappear via a supercritical Hopf bifurcation,
in consistent with the results found in Ref.~\cite{ref:shinomoto86}.
In the present study,
we further investigate the dependence of the collective phase sensitivity function on the noise intensity.
The typical shapes of the collective phase sensitivity function $\zeta(\Theta)$
near the onset of collective oscillations are displayed in Fig.~\ref{fig:4}(b).
The collective phase sensitivity function near the saddle-node bifurcation ($D = 0.10$)
satisfies $\zeta(\Theta) > 0$, which is referred to as type-I~\cite{ref:hansel95,ref:ermentrout96}.
That is, the collective phase is always advanced by positive weak impulses.
In contrast, the collective phase sensitivity function near the Hopf bifurcation ($D = 0.25$)
is type-II~\cite{ref:hansel95,ref:ermentrout96},
that is, $\zeta(\Theta)$ has negative parts as well as positive parts.

The above results are the same as those for the finite-dimensional dynamical system~\cite{ref:ermentrout96}.
However, note that they are the macroscopic collective characteristics of an infinite-dimensional dynamical system
that emerged from interactions of individual, microscopic elements.

As shown in this example of globally coupled active rotators,
a coupled dynamical system may have multiple routes
(such as saddle-node bifurcation or Hopf bifurcation)
to collective oscillations depending on the system parameters,
yielding different collective phase sensitivity functions.
It is known that in weakly interacting ordinary limit-cycle oscillators,
the difference in the type of phase sensitivity function
can lead to very different synchronization dynamics~\cite{ref:hansel95,ref:ermentrout96}.
Therefore, the present results imply that
when we consider populations of coupled dynamical elements undergoing collective oscillations,
the bifurcation of each population leading to the collective oscillations
can strongly influence the synchronization properties.

%%%%% section 4
\section{Concluding remarks} \label{sec:conclusion}

We developed a theory of collective phase description for globally coupled noisy active rotators
where macroscopic rhythms are generated by excitable elements.
By projecting the dynamics of the nonlinear Fokker-Planck equation
onto its stable limit-cycle orbit in the infinite-dimensional state space,
we derived a collective phase equation describing macroscopic rhythms of the system.
On the basis of our theory,
we analyzed the type of the collective phase sensitivity function near the onset of collective oscillations.
We obtained two distinct types of the collective phase sensitivity function;
they correspond to two different bifurcation routes to collective oscillations.

The theory we developed in this study can be considered a phase reduction method
for limit-cycle solutions to infinite-dimensional dynamical systems (see App.~\ref{sec:A}),
and includes our previous theory for phase oscillators as a special case (see App.~\ref{sec:B}).
Furthermore, the extension of this approach
to general systems of globally coupled dynamical elements
is briefly discussed in App.~\ref{sec:C}.

Finally, we remark the relation between the {\it phase}
and the breaking of continuously translational symmetry,
which is the core of the phase reduction theory~\cite{ref:kuramoto84}.
The phase of the ordinary limit-cycle solution
is associated with temporal translational symmetry breaking.
In Refs.~\cite{ref:kawamura07,ref:kawamura08,ref:kawamura10a},
the collective phase of the nonlinear Fokker-Planck equation of globally coupled phase oscillators
is assigned to a mixture of the spatial and temporal translational symmetry breakings,
where the term ``spatial'' refers to the variable $\phi$.
The formulation in Refs.~\cite{ref:kawamura07,ref:kawamura08,ref:kawamura10a}
is also essentially the same as the phase dynamics of wavefronts
in reaction-diffusion systems~\cite{ref:kuramoto84}.
However, as in the case of ordinary limit cycles,
spatial translational symmetry is not essential
(although it is typically assumed in the wavefront dynamics of reaction-diffusion systems)
for the phase description of limit-cycle solutions in infinite-dimensional dynamical systems.
In this paper, the collective phase of the nonlinear Fokker-Planck equation
is assigned only to the temporal translational symmetry breaking.
Therefore, the formulation is applicable to nonlinear Fokker-Planck equations
describing not only phase oscillators but also active rotators,
even though the latter case does not possess spatial translational symmetry.
A similar formulation for stable time-periodic solutions to reaction-diffusion systems
will be presented in Ref.~\cite{ref:nakao11}.

%%%%% acknowledgments
\begin{acknowledgments}
  The authors thank Kensuke Arai, Hiroshi Kori, and Naoki Masuda for fruitful discussions.
  % Tatsuo Yanagita % Shigeru Shinomoto
\end{acknowledgments}

\appendix

%%%%% section A
\section{Conventional phase reduction of limit-cycle oscillators} \label{sec:A}

We briefly review the conventional phase reduction method for coupled limit-cycle oscillators
~\cite{ref:kuramoto84,ref:hoppensteadt97,ref:izhikevich07}
to make a comparison to our present theory.
This will clarify that
the formulation in Sec.~\ref{sec:formulation} can be considered a phase reduction method
for limit-cycle solutions in infinite-dimensional dynamical systems.

%%% subsection
\subsection{Coupled limit-cycle oscillators}

We consider coupled limit-cycle oscillators described by the following model:
%%% eq
\begin{equation}
  \dot{\bd{X}}^{(\sigma)}\left( t \right)
  = \bd{F}\left( \bd{X}^{(\sigma)} \right)
  + \epsilon_p \bd{p}_\sigma\left( t \right)
  + \epsilon_g \bd{G}_{\sigma\tau}\left( \bd{X}^{(\sigma)}, \bd{X}^{(\tau)} \right),
  \label{eq:secA-model}
\end{equation}
for $(\sigma, \tau) = (1, 2)$ or $(2, 1)$,
where $\bd{X}^{(\sigma)}(t)$
is the $d$-dimensional state of the $\sigma$-th limit-cycle oscillator at time $t$.
The first term on the right-hand side represents the intrinsic dynamics of the limit-cycle oscillator,
the second term represents the external forcing,
and the last term represents the mutual coupling between the oscillators.
The dynamics of the isolated limit-cycle oscillator are given by $\dot{\bd{X}} = \bd{F}(\bd{X})$.
The external forcing is denoted by $\bd{p}_\sigma(t)$,
whose characteristic intensity is given by $\epsilon_p \geq 0$.
The mutual coupling function is
$\bd{G}_{\sigma\tau}(\bd{X}^{(\sigma)}, \bd{X}^{(\tau)})$,
whose characteristic intensity is $\epsilon_g \geq 0$.

%%% subsection
\subsection{Limit-cycle solution and its Floquet eigenvectors}

Let us assume $\epsilon_p = \epsilon_g = 0$ and focus on a single isolated oscillator.
The oscillator index $\sigma$ is dropped in this subsection.
The limit-cycle solution to Eq.~(\ref{eq:secA-model})
without external forcing or mutual coupling can be described by
%%% eq
\begin{equation}
  \bd{X}\left( t \right) = \bd{X}_0\left( \theta \right), \qquad
  \dot{\theta}(t) = \omega,
  \label{eq:secA-solution}
\end{equation}
where $\theta$ and $\omega$ are phase and frequency, respectively.
Inserting Eq.~(\ref{eq:secA-solution})
into Eq.~(\ref{eq:secA-model}) with $\epsilon_p = \epsilon_g = 0$,
we find that $\bd{X}_0(\theta)$ satisfies the following equation:
%%% eq
\begin{equation}
  \omega \frac{d}{d\theta} \bd{X}_0\left( \theta \right)
  = \bd{F}\left( \bd{X}_0\left( \theta \right) \right).
  \label{eq:secA-X0}
\end{equation}
Let $\bd{u}(\theta, t)$ represent a small disturbance to the limit-cycle solution;
let us then consider a slightly perturbed solution
%%% eq
\begin{equation}
  \bd{X}\left( t \right) = \bd{X}_0\left( \theta \right) + \bd{u}\left( \theta, t \right).
\end{equation}
Equation~(\ref{eq:secA-model}) with $\epsilon_p = \epsilon_g = 0$
is then linearized in $\bd{u}(\theta, t)$ as
%%% eq
\begin{equation}
  \frac{\partial}{\partial t} \bd{u}\left( \theta, t \right)
  = \hat{L}\left( \theta \right) \bd{u}\left( \theta, t \right),
  \label{eq:secA-linear}
\end{equation}
where the linear operator $\hat{L}(\theta)$ is given by
%%% eq
\begin{equation}
  \hat{L}\left( \theta \right) \bd{u}\left( \theta \right)
  = \left[ \hat{J}\left( \theta \right) - \omega \frac{\partial}{\partial \theta} \right]
  \bd{u}\left( \theta \right)
\end{equation}
with a Jacobi matrix
%%% eq
\begin{equation}
  \hat{J}\left( \theta \right)
  = \frac{\partial \bd{F}(\bd{X}_0(\theta))}{\partial \bd{X}_0(\theta)}.
\end{equation}
Defining the inner product as
%%% eq
\begin{equation}
  \Pd{ \bd{u}^\ast\left( \theta \right),\, \bd{u}\left( \theta \right) }
  = \frac{1}{2\pi} \int_0^{2\pi} d\theta \,
  \bd{u}^\ast\left( \theta \right) \cdot \bd{u}\left( \theta \right),
  \label{eq:secA-inner}
\end{equation}
we introduce an adjoint operator of $\hat{L}(\theta)$ by
%%% eq
\begin{equation}
  \Pd{ \bd{u}^\ast\left( \theta \right),\,
    \hat{L}\left( \theta \right) \bd{u}\left( \theta \right) }
  = \Pd{ \hat{L}^\ast\left( \theta \right) \bd{u}^\ast\left( \theta \right),\,
    \bd{u}\left( \theta \right) }.
\end{equation}
The adjoint operator $\hat{L}^\ast(\theta)$ is explicitly given by
%%% eq
\begin{equation}
  \hat{L}^\ast\left( \theta \right) \bd{u}^\ast\left( \theta \right)
  = \left[ \hat{J}^\ast\left( \theta \right) + \omega \frac{\partial}{\partial \theta} \right]
  \bd{u}^\ast\left( \theta \right)
\end{equation}
with the transposed matrix of the Jacobi matrix
%%% eq
\begin{equation}
  \hat{J}^\ast(\theta) = \hat{J}(\theta)^{\rm T}.
\end{equation}
We use the Floquet eigenvectors of the periodic linear system Eq.~(\ref{eq:secA-linear}) and its adjoint system
associated with the zero eigenvalue, i.e.,
%%% eq
\begin{align}
  \hat{L}\left( \theta \right) \bd{u}_0\left( \theta \right)
  = \left[ \hat{J}\left( \theta \right) - \omega \frac{d}{d\theta} \right]
  \bd{u}_0\left( \theta \right)
  &= 0, \\
  \hat{L}^\ast\left( \theta \right) \bd{u}_0^\ast\left( \theta \right)
  = \left[ \hat{J}^\ast\left( \theta \right) + \omega \frac{d}{d\theta} \right]
  \bd{u}_0^\ast\left( \theta \right)
  &= 0. \label{eq:secA-leftzero}
\end{align}
Note that the right zero eigenvector $\bd{u}_0(\theta)$ can be chosen as
%%% eq
\begin{equation}
  \bd{u}_0\left( \theta \right) = \frac{d}{d\theta} \bd{X}_0\left( \theta \right),
  \label{eq:secA-u0}
\end{equation}
which is confirmed by differentiating Eq.~(\ref{eq:secA-X0}) with respect to $\theta$.
Using the inner product~(\ref{eq:secA-inner}) with the right zero eigenvector~(\ref{eq:secA-u0}),
the left zero eigenvector $\bd{u}_0^\ast(\theta)$ is normalized as
%%% eq
\begin{equation}
  \Pd{ \bd{u}_0^\ast\left( \theta \right),\, \bd{u}_0\left( \theta \right) }
  = \frac{1}{2\pi} \int_0^{2\pi} d\theta \,
  \bd{u}_0^\ast\left( \theta \right) \cdot \bd{u}_0\left( \theta \right)
  = 1.
\end{equation}
As in the infinite-dimensional case treated in the main text,
the following equation holds~\cite{ref:hoppensteadt97}:
%%% eq
\begin{align}
  \frac{d}{d\theta}
  \Bigl[ \bd{u}_0^\ast\left( \theta \right) \cdot \bd{u}_0\left( \theta \right) \Bigr]
  &= \bd{u}_0^\ast\left( \theta \right) \cdot \frac{d\bd{u}_0\left( \theta \right)}{d\theta}
  + \frac{d\bd{u}_0^\ast\left( \theta \right)}{d\theta} \cdot \bd{u}_0\left( \theta \right)
  \nonumber \\
  &= \frac{1}{\omega} \left[
    \bd{u}_0^\ast\left( \theta \right) \cdot \hat{J}\left( \theta \right) \bd{u}_0\left( \theta \right)
    - \hat{J}^\ast\left( \theta \right) \bd{u}_0^\ast\left( \theta \right) \cdot \bd{u}_0\left( \theta \right) \right]
  \nonumber \\
  &= 0.
  \label{eq:secA-dtheta}
\end{align}
Therefore, the normalization condition is satisfied separately for each $\theta$ as follows
%%% eq
\begin{equation}
  \bd{u}_0^\ast\left( \theta \right) \cdot \bd{u}_0\left( \theta \right) = 1.
\end{equation}
The left zero eigenvector is the so-called phase sensitivity function~\cite{ref:kuramoto84}, i.e.,
%%% eq
\begin{equation}
  \bd{u}_0^\ast\left( \theta \right) = \bd{Z}\left( \theta \right).
\end{equation}
%%

%%% subsection
\subsection{Phase reduction and the normal form}

Now, we introduce the effects of weak external forcing and mutual coupling,
and treat the second and last terms in Eq.~(\ref{eq:secA-model}) as perturbations.
Using the phase reduction method~\cite{ref:kuramoto84},
we can derive a phase equation from Eq.~(\ref{eq:secA-model}) as follows:
%%% eq
\begin{equation}
  \dot{\theta}^{(\sigma)}\left( t \right) = \omega
  + \epsilon_p \bd{Z}\left( \theta^{(\sigma)} \right) \cdot \bd{p}_\sigma\left( t \right)
  + \epsilon_g \bd{Z}\left( \theta^{(\sigma)} \right) \cdot
  \bd{G}_{\sigma\tau}\left( \bd{X}_0\left( \theta^{(\sigma)} \right), \bd{X}_0\left( \theta^{(\tau)} \right) \right).
\end{equation}
Applying the near-identity transformation~\cite{ref:kuramoto84,ref:hoppensteadt97,ref:kori09},
we can obtain the following simplified phase equation:
%%% eq
\begin{equation}
  \dot{\theta}^{(\sigma)}\left( t \right) = \omega
  + \epsilon_p \bd{Z}\left( \theta^{(\sigma)} \right) \cdot \bd{p}_\sigma\left( t \right)
  + \epsilon_g \Gamma_{\sigma\tau}\left( \theta^{(\sigma)} - \theta^{(\tau)} \right),
\end{equation}
where the phase coupling function is given by
%%% eq
\begin{equation}
  \Gamma_{\sigma\tau}\left( \theta \right)
  = \frac{1}{2\pi} \int_0^{2\pi} d\lambda\, \bd{Z}\left( \lambda + \theta \right)
  \cdot \bd{G}_{\sigma\tau}\left( \bd{X}_0( \lambda + \theta ), \bd{X}_0( \lambda ) \right).
\end{equation}
%%

%%% subsection
\subsection{Adjoint method for the left zero eigenvector}

From Eq.~(\ref{eq:secA-leftzero}),
the left zero eigenvector $\bd{u}_0^\ast(\theta)$ satisfies
%%% eq
\begin{equation}
  \omega \frac{d}{d\theta} \bd{u}_0^\ast\left( \theta \right)
  = - \hat{J}^\ast\left( \theta \right) \bd{u}_0^\ast\left( \theta \right),
\end{equation}
which can be transformed into
%%% eq
\begin{equation}
  \frac{d}{ds} \bd{u}_0^\ast\left( -\omega s \right)
  = \hat{J}^\ast\left( -\omega s \right) \bd{u}_0^\ast\left( -\omega s \right)
  \label{eq:secA-adjoint}
\end{equation}
with $\theta = -\omega s$.
As is well known~\cite{ref:ermentrout96,ref:hoppensteadt97,ref:izhikevich07},
the relaxation method using Eq.~(\ref{eq:secA-adjoint})
is a convenient method for obtaining the phase sensitivity function $\bd{Z}(\theta)$ numerically.

%%% subsection
\subsection{Remarks on the relation to the conventional phase reduction of ordinary limit cycles}

Now, we see that the formulation in Sec.~\ref{sec:formulation}
clearly corresponds to that in this appendix.
In particular, the following relations are important:
%%% eq
\begin{align}
  f_0\left( \phi, \Theta \right)
  &\leftrightarrow \bd{X}_0(\theta), \\
  \int_0^{2\pi} d\phi\, u_0^\ast\left( \phi, \Theta \right) u_0\left( \phi, \Theta \right)
  &\leftrightarrow \bd{u}_0^\ast(\theta) \cdot \bd{u}_0(\theta).
\end{align}
Namely, the formulation in Sec.~\ref{sec:formulation}
is considered a phase reduction of the limit-cycle solution
to the infinite-dimensional dynamical system
that is continuously parameterized by the variable $\phi$.

%%%%% section B
\section{Phase oscillators with translational symmetry} \label{sec:B}

Here we clarify the relation between the formulation developed in Sec.~\ref{sec:formulation}
and our previous formulation for globally coupled phase oscillators
~\cite{ref:kawamura07,ref:kawamura08,ref:kawamura10a}.
Phase oscillators can be described as a special case of active rotators as follows:
%%% eq
\begin{align}
  v\left( \phi_j \right)
  &\to \omega, \\
  \Gamma\left( \phi_j, \phi_k \right)
  &\to \Gamma\left( \phi_j - \phi_k \right).
\end{align}
In this case,
%the nonlinear Fokker-Planck equation~(\ref{eq:nfp}) with $\epsilon_p = \epsilon_g = 0$
the nonlinear Fokker-Planck equation~(\ref{eq:single-nfp})
has a translational symmetry with respect to $\phi$.
Reflecting this symmetry,
the collectively oscillating solution and other associated functions have the following properties:
%%% eq
\begin{align}
  & f_0\left( \phi, \Theta \right)
  = f_0\left( \phi - \Theta, 0 \right)
  \equiv f_0\left( \varphi \right), \\
  & u_0\left( \phi, \Theta \right)
  = u_0\left( \phi - \Theta, 0 \right)
  \equiv u_0\left( \varphi \right)
  = -\frac{d}{d\varphi}f_0\left( \varphi \right), \label{eq:secB-u0} \\
  & u_0^\ast\left( \phi, \Theta \right)
  = u_0^\ast\left( \phi - \Theta, 0 \right)
  \equiv u_0^\ast\left( \varphi \right), \\
  & k_0\left( \phi, \Theta \right)
  = k_0\left( \phi - \Theta, 0 \right)
  \equiv k_0\left( \varphi \right)
  = f_0\left( \varphi \right) \frac{d}{d\varphi}u_0^{\ast}\left( \varphi \right).
\end{align}
In addition, the kernel function $k_0(\varphi)$ is normalized as
%%% eq
\begin{equation}
  \int_0^{2\pi} d\varphi\, k_0\left( \varphi \right)
  = \int_0^{2\pi} d\varphi\, f_0\left( \varphi \right) \frac{d}{d\varphi}u_0^{\ast}\left( \varphi \right)
  = \int_0^{2\pi} d\varphi\, u_0^{\ast}\left( \varphi \right) u_0\left( \varphi \right) 
  = 1.
\end{equation}
The collective phase sensitivity function $\zeta(\Theta)$ is given by
%%% eq
\begin{equation}
  \zeta\left( \Theta \right)
  = \int_0^{2\pi} d\phi\, Z\left( \phi \right) k_0\left( \phi - \Theta \right)
  = \int_0^{2\pi} d\varphi\, Z\left( \varphi + \Theta \right) k_0\left( \varphi \right),
  \label{eq:secB-zeta}
\end{equation}
whose $\Theta$-dependence exists only when the phase sensitivity function $Z(\phi)$ is not constant.
Furthermore, the external coupling function $\Gamma_{\sigma\tau}$ of the phase oscillators
can be written in the following form:
%%% eq
\begin{equation}
  \Gamma_{\sigma\tau}\left( \phi_j^{(\sigma)}, \phi_k^{(\tau)} \right)
  \to \Gamma_{\sigma\tau}\left( \phi_j^{(\sigma)} - \phi_k^{(\tau)} \right).
\end{equation}
Again, reflecting the translational symmetry,
the effective phase coupling function $\digamma_{\sigma\tau}$
depends only on the collective phase difference
and coincides with the collective phase coupling function $\gamma_{\sigma\tau}$ as follows:
%%% eq
\begin{equation}
  \digamma_{\sigma\tau}\left( \Theta^{(\sigma)}, \Theta^{(\tau)} \right)
  = \digamma_{\sigma\tau}\left( \Theta^{(\sigma)} - \Theta^{(\tau)}, 0 \right)
  \equiv \digamma_{\sigma\tau}\left( \Theta^{(\sigma)} - \Theta^{(\tau)} \right)
  = \gamma_{\sigma\tau}\left( \Theta^{(\sigma)} - \Theta^{(\tau)} \right),
\end{equation}
or, explicitly,
%%% eq
\begin{align}
  \gamma_{\sigma\tau}\left( \Theta^{(\sigma)} - \Theta^{(\tau)} \right)
  &= \int_0^{2\pi} d\phi \int_0^{2\pi} d\phi'\, \Gamma_{\sigma\tau}\left( \phi - \phi' \right)
  k_0\left( \phi - \Theta^{(\sigma)} \right) f_0\left( \phi' - \Theta^{(\tau)} \right) \nonumber \\
  &= \int_0^{2\pi} d\varphi \int_0^{2\pi} d\varphi'\,
  \Gamma_{\sigma\tau}\left( \varphi - \varphi' + \Theta^{(\sigma)} - \Theta^{(\tau)} \right)
  k_0\left( \varphi \right) f_0\left( \varphi' \right).
  \label{eq:secB-gamma}
\end{align}
Here, we should note that when the external coupling function $\Gamma_{\sigma\tau}$ is sinusoidal,
the collective phase coupling function $\gamma_{\sigma\tau}$ also has a sinusoidal form.
The above collective phase sensitivity and coupling functions for the phase oscillators,
i.e., Eq.~(\ref{eq:secB-zeta}) and Eq.~(\ref{eq:secB-gamma}),
are equivalent to the formulae that we derived in Refs.~\cite{ref:kawamura08,ref:kawamura10a}.
Precisely speaking, the above system of notation is slightly different from
the previous one~\cite{ref:kawamura07,ref:kawamura08,ref:kawamura10a} as follows:
%%% eq
\begin{align}
  f_0\left( \varphi \right)
  &= \tilde{f}_0\left( \varphi \right), \\
  u_0\left( \varphi \right)
  &= -\tilde{u}_0\left( \varphi \right), \\
  u_0^\ast\left( \varphi \right)
  &= -\tilde{u}_0^\ast\left( \varphi \right), \\
  k_0^\ast\left( \varphi \right)
  = f_0\left( \varphi \right) \frac{d}{d\varphi}u_0^{\ast}\left( \varphi \right)
  &= -\tilde{f}_0\left( \varphi \right) \frac{d}{d\varphi}\tilde{u}_0^{\ast}\left( \varphi \right)
  = \tilde{k}_0^\ast\left( \varphi \right).
\end{align}
Here, the previous notations of Refs.~\cite{ref:kawamura07,ref:kawamura08,ref:kawamura10a} are denoted by
$\tilde{f}_0(\varphi)$, $\tilde{u}_0(\varphi)$, $\tilde{u}_0^\ast(\varphi)$, and $\tilde{k}_0(\varphi)$.
These differences, which are not essential,
come from the fact that the right zero eigenfunction was chosen as
$\tilde{u}_0(\varphi) = d\tilde{f}_0(\varphi)/d\varphi$,
in contrast to Eq.~(\ref{eq:secB-u0}), i.e., $u_0(\varphi) = -df_0(\varphi)/d\varphi$.

\begin{figure*}
  \begin{center}
    \includegraphics[width=0.9\hsize,clip]{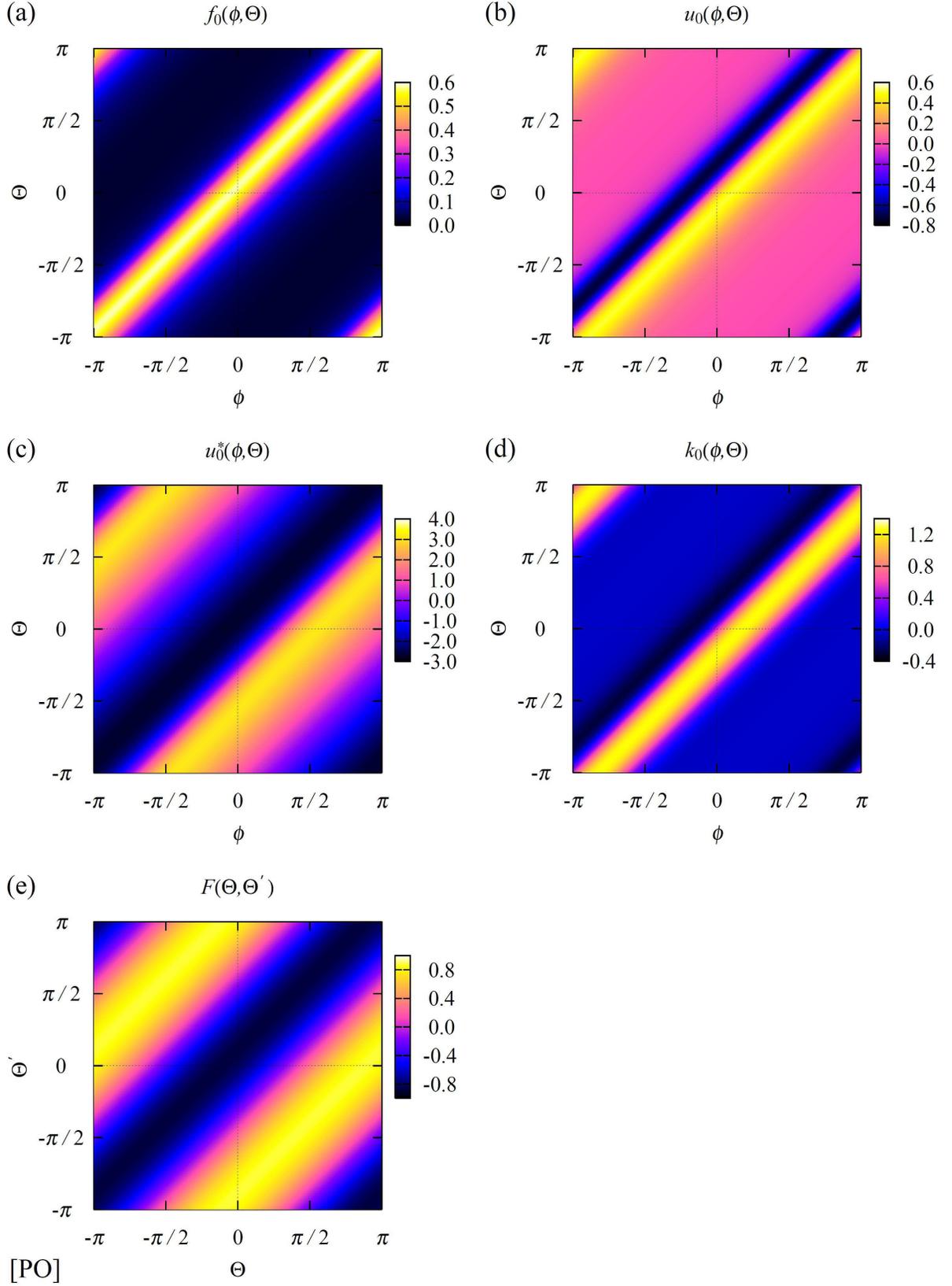}
    \caption{(Color online)
      Globally coupled phase oscillators (PO)
      corresponding to the translationally symmetric case.
      (a) Collectively oscillating solution $f_0(\phi, \Theta)$.
      (b) Right zero eigenfunction $u_0(\phi, \Theta)$.
      (c) Left zero eigenfunction $u_0^\ast(\phi, \Theta)$.
      (d) Kernel function $k_0(\phi, \Theta)$.
      (e) Effective phase coupling function $\digamma(\Theta, \Theta')$.
      Parameters are $\alpha = 3\pi/8$ and $D = D_{\rm c}/2 = \cos(\alpha)/4$.
    }
    \label{fig:A1}
  \end{center}
  %%%%% parameter
  %% 
  %% v(\phi) = \omega,
  %% \Gamma(\phi - \phi') = -\sin(\phi - \phi' + \alpha).
  %% 
  %% \alpha = 3\pi/8, D = D_c/2 = \cos(\alpha)/4.
  %% 
  %%%%%
\end{figure*}

\begin{figure*}
  \begin{center}
    \includegraphics[width=0.9\hsize,clip]{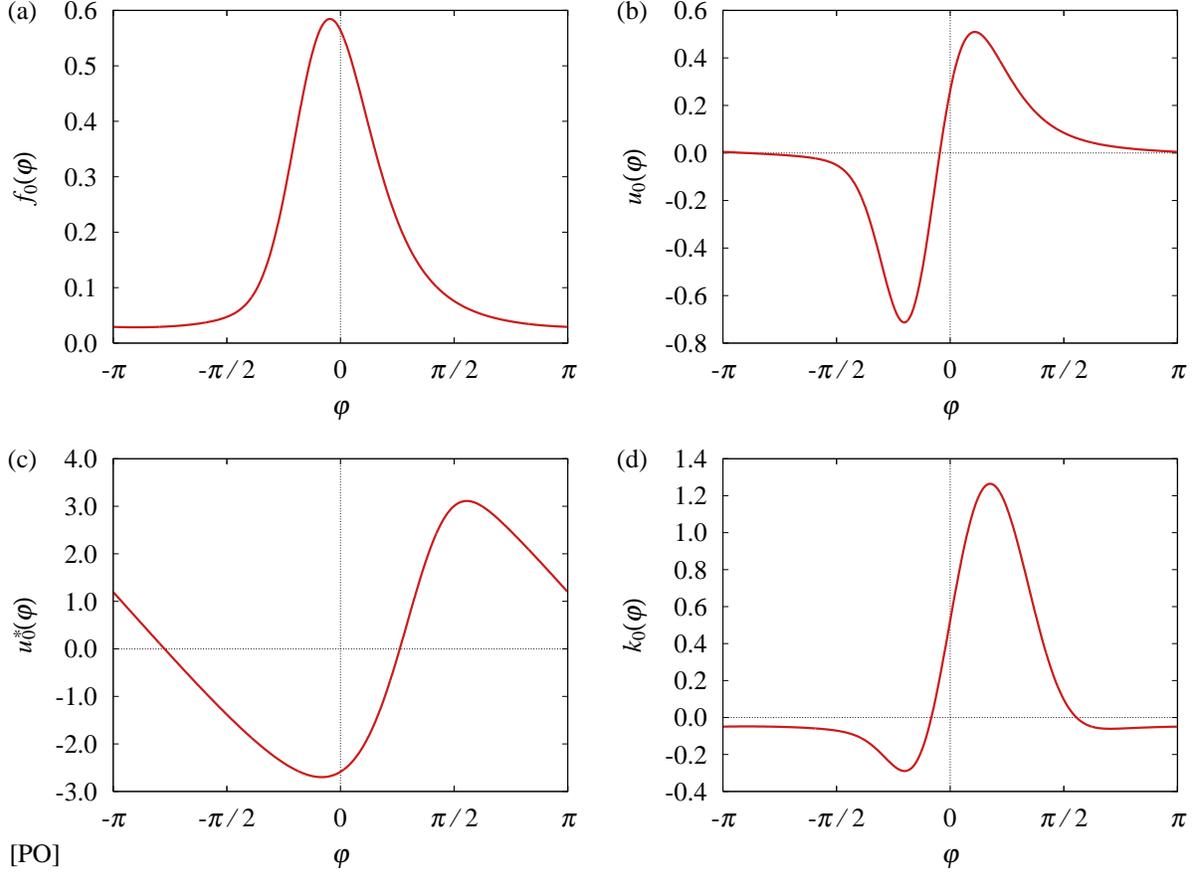}
    \caption{(Color online)
      Globally coupled phase oscillators (PO).
      (a) Collectively oscillating solution $f_0(\varphi)$.
      (b) Right zero eigenfunction $u_0(\varphi)$.
      (c) Left zero eigenfunction $u_0^\ast(\varphi)$.
      (d) Kernel function $k_0(\varphi)$.
      Parameters are $\alpha = 3\pi/8$ and $D = D_{\rm c}/2 = \cos(\alpha)/4$.
    }
    \label{fig:A2}
  \end{center}
  %%%%% parameter
  %% 
  %% see Fig.A1.
  %% 
  %%%%%
\end{figure*}

Let us consider the following simple model:
%%% eq
\begin{equation}
  \Gamma\left( \phi \right)
  = \Gamma_{\sigma\tau}\left( \phi \right)
  = -\sin\left( \phi + \alpha \right), \qquad
  | \alpha | < \frac{\pi}{2}.
\end{equation}
In this case, collective oscillations exist under the condition $D < D_{\rm c} = \cos(\alpha)/2$
~\cite{ref:kuramoto84,ref:kawamura10a}.
The collectively oscillating solution $f_0(\phi,\Theta)$
and other associated functions, i.e., $u_0(\phi,\Theta)$, $u_0^\ast(\phi,\Theta)$,
$k_0(\phi,\Theta)$, and $\digamma(\Theta,\Theta')$, are shown in Fig.~\ref{fig:A1},
where the parameters are set to $\alpha = 3\pi/8$ and $D = D_{\rm c}/2 = \cos(\alpha)/4$.
All functions are translationally symmetric in sharp contrast to Fig.~\ref{fig:2}.
Using the translational symmetry,
Figs.~\ref{fig:A1}(a)(b)(c)(d) can be compressed into Fig.~\ref{fig:A2},
which is essentially the same as the results in Ref.~\cite{ref:kawamura10a}.

%%%%% section C
\section{Extension to general systems of globally coupled elements} \label{sec:C}

The formulation in Sec.~\ref{sec:formulation} can be straightforwardly
extended to general systems of globally coupled noisy elements,
whose outline is sketched below.
We can consider general dynamical elements such as excitable, oscillatory, or chaotic units,
as long as the system exhibits collective oscillations described by a time-periodic solution.
We consider a group of globally coupled general dynamical elements
subjected to independent white Gaussian noise described by
%%% eq
\begin{equation}
  \dot{\bd{X}}_j\left( t \right)
  = \bd{F}\left( \bd{X}_j \right)
  + \frac{1}{N} \sum_{k=1}^N \bd{G}\left( \bd{X}_j, \bd{X}_k \right)
  + \sqrt{D}\, \bd{\xi}_j\left( t \right),
\end{equation}
which can be transformed into the following nonlinear Fokker-Planck equation:
%%% eq
\begin{equation}
  \frac{\partial}{\partial t} f\left( \bd{X}, t \right) =
  -\frac{\partial}{\partial \bd{X}} \cdot \left[ \left\{
    \bd{F}\left( \bd{X} \right) + \int d\bd{X}'\, \bd{G}\left( \bd{X}, \bd{X}' \right) f\left( \bd{X}', t \right)
    \right\} f\left( \bd{X}, t \right) \right]
  + D \left( \frac{\partial}{\partial \bd{X}} \cdot \frac{\partial}{\partial \bd{X}} \right) f\left( \bd{X}, t \right).
\end{equation}
Therefore, a collectively oscillating solution can be described by
%%% eq
\begin{equation}
  f\left( \bd{X}, t \right) = f_0\left( \bd{X}, \Theta \right), \qquad
  \dot{\Theta}(t) = \Omega,
\end{equation}
whose right zero eigenfunction can be chosen as
%%% eq
\begin{equation}
  u_0\left( \bd{X}, \Theta \right) = \frac{\partial}{\partial \Theta} f_0\left( \bd{X}, \Theta \right),
\end{equation}
and whose left zero eigenfunction is normalized as
%%% eq
\begin{equation}
  \int d\bd{X}\, u_0^\ast\left( \bd{X}, \Theta \right) u_0\left( \bd{X}, \Theta \right) = 1.
\end{equation}
It is clear that collective phase description for general systems
of globally coupled noisy elements exhibiting macroscopic rhythms
can be formulated, in principle, in the same way.

%%%%% references

\clearpage

%%%%% figures

%%% fig.1
%%

%%% fig.2
%%

%%% fig.3
%%

%%% fig.4
%%

%%% fig.A1
%%

%%% fig.A2

\end{document}